\documentclass[aps,10pt,final,
notitlepage, oneside, twocolumn, nobibnotes, nofootinbib,
superscriptaddress, showpacs, centertags, showkeys, assume]{revtex4}

\usepackage{graphicx}
\usepackage{amsmath}

\begin{document}

\title{Sum rule for photon target}

\author{E. Barto\v s}\affiliation{Inst.\ of Physics,
Slovak Academy of\ Sciences, D\'ubravsk\'a cesta 9, 845 11
Bratislava, Slovak Republic}
\author{S.~Dubni\v{c}ka} \affiliation{Inst.\ of Physics,
Slovak Academy of\ Sciences, D\'ubravsk\'a cesta 9, 845 11
Bratislava, Slovak Republic}
\author{A.Z.~Dubni\v ckov\'a}
\affiliation{Dept.\ of Theor. Physics, Comenius University, 842 48
Bratislava, Slovak Republic}
\author{E.~A.~Kuraev} \affiliation{Bogol'ubov\ Laboratory of Theoretical
Physics, JINR, Dubna, Russia}

\begin{abstract}
The amplitude of zero angle scattering of electron on photon in
the 3-rd  QED order of fine structure constant with
$\gamma^*\gamma$ intermediate state  converting into
quark--antiquark is considered. Utilizing analytic properties of
elastic photon--photon scattering amplitude an explicit expression
for differential cross- section of quark--antiquark pair
production at electron-photon collision in peripheral kinematics
is derived apparently. Limiting case of small transferred momenta
with an application of the Weizs\"acker-Williams like relation
gives the sum rule for photon target, bringing into the relation
the sum of ratios of the four power of the quark charges to
squared quark masses with integral over the total $\gamma\gamma
\to 2 jets$ cross--section.
\end{abstract}

\maketitle

Sum rules connecting the high energy asymptotic cross-sections of
peripheral processes of QED type, which can be studied in colliding
electron-positron beams with the fermion form factors (in particular
with the slope of Dirac form factor) have been found in 1974
\cite{Kuraev:1974jd}. Applications to more complicated two-loop
level QED processes have been investigated in the Appendix of the
review paper \cite{Baier:1980kx}. In a set of the papers
\cite{BDK}-\cite{dub07} some applications to baryon, deuteron and
meson form factors were considered, where connection of these
electromagnetic form factors with $Q^2$ - dependent differential
cross-section of deep inelastic electron-hadron scattering in the
peripheral kinematics was investigated.

In this paper we consider the scattering of electron on photon
target with creation of 2 jets in the fragmentation region of the
photon. Such kind of problems can be searched at photon-electron
colliders constructed on the base of linear electron-positron
colliders.
\begin{figure*}[htb]
\centerline{\includegraphics[width=0.15\textwidth]{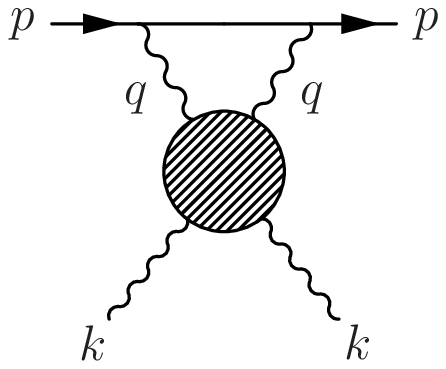} \qquad
\includegraphics[width=0.45\textwidth]{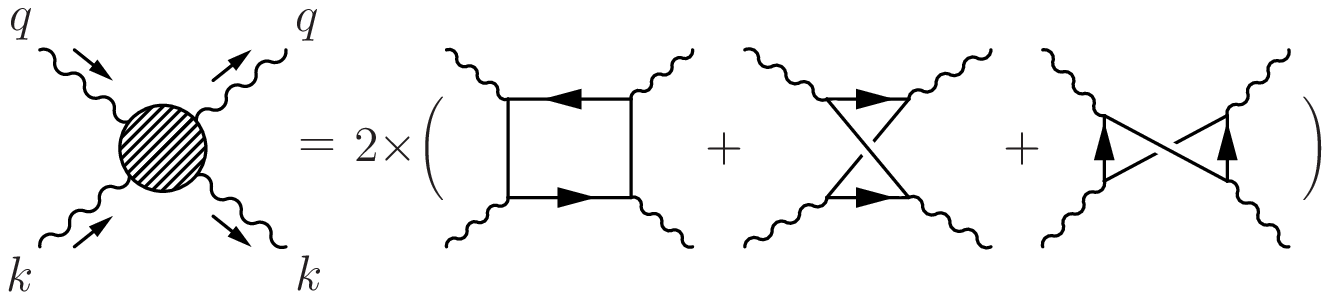}}
\caption{Feynman diagram of $e\gamma\to e\gamma$ scattering with LBL
mechanism to be realized by quark-loops}
\end{figure*}

Let us consider the two photon exchange electron-photon zero angle
scattering amplitude of the process
\begin{equation} \label{eq:proc}
e(p,\lambda)+\gamma(k,\varepsilon)\to
e(p,\lambda)+\gamma(k,\varepsilon),
\end{equation}
in two-loop ($\alpha^3$) approximation as presented in Fig. 1,
with $p^2=m_{e}^2$, $k^2=0$ and assuming that the total energy
squared of the process (\ref{eq:proc}) $s=2p.k\gg m_{e}^2$.

   In an explicit calculation of the corresponding amplitude and
a derivation of the sum rule under consideration the Sudakov's
decomposition of the virtual photon transferred momentum
\begin{eqnarray}\label{eq:sud}
q= \alpha \tilde{p}+\beta k+q_{\bot},\quad q_{\bot}=(0,0,\vec
q),\quad q_{\bot}^2=-\vec q^2, \\
\tilde{p}q_{\bot}=kq_{\bot}=0\nonumber
\end{eqnarray}
into light-like vector $\tilde{p}=p-kp^2/s$, \quad $\tilde{p}^2=0$
and the real photon fourmomentum $k$ is suitable. The total energy
squared variable $s_1$ of the photon-photon scattering subprocess
is then
\begin{gather}
s_1=(q+k)^2= \alpha s - \vec q^2, \quad
 d^4q=d^2\vec q \frac{ds_1}{2}d\beta.
\end{gather}
Averaging over the initial electron and photon spin states
(initial and final spin states are supposed to coincide) one can
write down the amplitude of the process (\ref{eq:proc}) in the
following form
\begin{gather} \label{eq:ampl}
A^{e\gamma\to e\gamma}(s,t=0)=\\ \nonumber
=s\frac{\alpha}{4\pi^2}\int\frac{d^2\vec{q}}{(q^2)^2}d s_1
\sum_\varepsilon
A^{\gamma\gamma\to\gamma\gamma}_{\mu\nu\alpha\beta}\frac{p^\mu
p^\nu \varepsilon^{\alpha} \varepsilon^{*\beta}}{s^2},
\end{gather}
where the light-cone projection of the light-by-light (LBL)
scattering tensor is the amplitude of $\gamma\gamma \to
\gamma\gamma$ process and takes the form
\begin{gather}\label{eq:a1}
A^{\gamma\gamma\to\gamma\gamma}(s_1,\vec q)=
A^{\gamma\gamma\to\gamma\gamma}_{\mu\nu\alpha\beta}\frac{p^\mu
p^\nu \varepsilon^{\alpha} \varepsilon^{*\beta}}{s^2}=\\ \nonumber
=-\frac{8\alpha^2}{\pi^2}N_cQ_q^4\int
d^4q_{-}\Big[\frac{S_1}{D_1}+\frac{S_2}{D_2}+\frac{S_3}{D_3}\Big ]
\end{gather}
with
\begin{widetext}
\begin{gather}
\frac{S_1}{D_1}=\frac{(1/4)Tr\hat{p}(\hat{q}_-+m_q)\hat{p}(\hat{q}_--\hat{q}+m_q)\hat{\varepsilon}^*
(\hat{q}_--\hat{q}+\hat{k}+m_q)\hat{\varepsilon}(\hat{q}_--\hat{q}+m_q)}{(q_-^2-m^2_q)((q_--q)^2-m^2_q)^2
((q_--q+k)^2-m^2_q)},\\
\frac{S_2}{D_2}=\frac{(1/4)Tr\hat{p}(\hat{q}_-+m_q)\hat{p}(\hat{q}_--\hat{q}+m_q)\hat{\varepsilon}^
(\hat{q}_--\hat{q}-\hat{k}+m_q)\hat{\varepsilon}^*(\hat{q}_--\hat{q}+m_q)}{(q_-^2-m^2_q)((q_--q)^2-m^2_q)^2
((q_--q-k)^2-m_q^2)},\\
\frac{S_3}{D_3}=\frac{(1/4)Tr\hat{p}(\hat{q}_-+m_q)\hat{\varepsilon}(\hat{q}_--\hat{k}+m_q)\hat{p}
(\hat{q}_-+\hat{q}-\hat{k}+m_q)\hat{\varepsilon}^*(\hat{q}_-+\hat{q}+m_q)}{(q_-^2-m^2_q)((q_-+q)^2-m^2_q)
((q_-+q+k)^2-m^2_q)((q_--k)^2-m^2_q)}.
\end{gather}
\end{widetext}
where $q_{-}$ means the quark four-momentum in the quark loop of
the process $\gamma\gamma\to\gamma\gamma$,
\begin{figure}[htb]
\centerline{\includegraphics[width=0.4\textwidth]{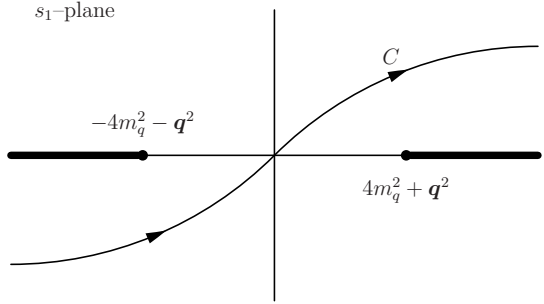}}\caption{The
path $C$ of an integration in (\ref{eq:a2})}
\end{figure}
$N_c$ is the number of colours in QCD and $Q_q$ is the charge of
the quark $q$ in electron charge units.

Regularization of LBL tensor is implied to provide the gauge
invariance, which consists in removing some constant symmetrical
tensor and the latter has no influence on the final results.

Now, taking a derivative of the relation (\ref{eq:a1}) according
to $d^2\vec{q}$ and investigating the analytic properties of the
obtained expression in $s_1$ -plane one gets the configuration as
presented in Fig. 2, where also the path $C$ of the integral
expression

\begin{equation}\label{eq:a2}
I=\int_Cds_1\frac{dA^{\gamma\gamma \to
\gamma\gamma}(s_1,\vec{q})}{d^2\vec{q}}
\end{equation}
is drawn. When the integration contour is closed to the right (on
s-channel cut) and to the left (on the $u$-channel cut) one comes to
the relation
\begin{eqnarray}\label{a3}
\int_{-4m_q^2-\vec{q}^2}^{-\infty}ds_1\Delta_u
\frac{{dA}^{\gamma\gamma
\to \gamma\gamma}(s_1,\vec{q})}{d^2\vec{q}}{|_{left}}=\\
\nonumber =\int_{4m_q^2+\vec{q}^2}^\infty ds_1\Delta_s
\frac{{dA}^{\gamma\gamma \to
\gamma\gamma}(s_1,\vec{q})}{d^2\vec{q}}{|_{right}},
\end{eqnarray}
where the right s-channel discontinuity by means of the equation
(\ref{eq:ampl}) is related (due to optical theorem in a
differential form)
\begin{equation}\label{a4}
\Delta_s\frac{dA^{e\gamma \to e\gamma}(s,0)}{d^2\vec{q}}=
2s\frac{d\sigma^{e\gamma\to e q\bar{q}}}{d^2\vec{q}},
\end{equation}
to the $Q^2$=$\vec{q}^2$=$-q^2$ dependent differential
cross-section of $q\bar q$ pair creation by electron on photon, to
be well known in the framework of QED \cite{KEL} for $l^+l^-$ pair
creation
\begin{eqnarray}\label{a6}
&&\frac{4\alpha^3}{3(q^2)^2}f(\frac{\vec{q}^2}{m^2_q})N_cQ_q^4=
\frac{d\sigma^{e\gamma \to e q\bar{q}}}{d\vec{q}^2},\\
&&f(\frac{\vec{q}^2}{m^2_q})=(\vec{q}^2-m^2_q)J+1,\nonumber \\
&&J=\frac{4}{\sqrt{\vec{q}^2(\vec{q}^2+4m^2_q)}}
\ln[\sqrt{\vec{q}^2/(4m^2_q)}+\sqrt{1+\vec{q}^2/(4m^2_q)}].\nonumber
\end{eqnarray}
But the right hand cut concerns of two real quark production for
$s_1> 4 m_q^2$, which is associated with 2 jets production.

The left-hand cut contribution has the same form as in QED case
with constituent quark masses and as a result one obtains
\begin{equation} \label{eq:sum}
\frac{4\alpha^3}{3(\vec{q}^2)^2}N_c\sum_q
Q_q^4f(\frac{\vec{q}^2}{m_q^2})=\frac{d\sigma^{e\gamma\to e2
jets}}{d\vec{q}^2}.
\end{equation}
Finally, for the case of small $\vec{q}^2$ and applying the
Weizs\"acker-Williams like relation one comes to the sum rule for
photon target as follows
\begin{equation}\label{a8}
\frac{14}{3}\sum_q\frac{Q_q^4}{m_q^2}=
\frac{1}{\pi\alpha^2}\int\limits_{4m^2_q}^{\infty} \frac{d
s_1}{s_1} \sigma_{tot}^{\gamma\gamma\to 2{jet}}(s_1).
\end{equation}

The quantity $\sigma_{tot}^{\gamma\gamma\to 2 jets}(s_1)$ is
assumed to degrease with increased values of $s_1$. It corresponds
to the events in $\gamma\gamma$ collisions with creation of two
jets, which are not separated by rapidity gaps and for which till
the present days there is no experimental information. The latter
complicates a verification of the sum rule (\ref{a8}).

An evaluation of the left-hand side with the constituent quark
masses \cite{NVY}  $m_u=m_d=280MeV$ and  $m_s=405 MeV$ (the
contributions of heavy quarks $c, b$ and $t$ give negligible
contributions and can be disregarded) gives just $5 mb$.

The saturation of the right-hand side of the photon sum rule
(\ref{a8}) with an utilization of the data on
$\sigma_{tot}^{\gamma\gamma\to X}(s_1)$ given by Review of
Particle Physics \cite{Yao} on the level of $5 mb$ is achieved
with the upper bound of the corresponding integral to be $2-3
GeV^2$. Unfortunately, the used data are charged by rather large
uncertainties and in order to achieve more reliable verification
of the sum rule (\ref{a8}) the data on
$\sigma_{tot}^{\gamma\gamma\to 2 jets}(s_1)$ are highly desirable.

\section{Acknowledgements}

The work was partly supported by Slovak Grant Agency for Sciences
VEGA, Grant No. 2/7116/2007 (E.B., S.D. and A.Z.D.). E.A. Kuraev
would like to thank Institute of Physics SAS for warm hospitality
and Slovak purposed project at JINR for financial support. All
authors thank Yu.M. Bystritskiy and M. Se\v cansk\'y for their
interest and for discussions at the early stage of this work.


\begin{thebibliography}{99}

\bibitem{Kuraev:1974jd}
  E.~A.~Kuraev, L.~N.~Lipatov and N.~P.~Merenkov,
  Yad.\ Fiz.\  {\bf 18} (1973) 1075.

\bibitem{Baier:1980kx}
  V.~N.~Baier, E.~A.~Kuraev, V.~S.~Fadin and V.~A.~Khoze,
  Phys.\ Rept.\  {\bf 78} (1981) 293.

\bibitem{BDK}
  E.~Barto\v{s}, S.~Dubni\v cka, E.~A.~Kuraev,
 Phys.\ Rev. D {\bf 70} (2004) 117901.

\bibitem{Kuraev:2006ym}
  E.~A.~Kuraev, M.~Secansky and E.~Tomasi-Gustafsson,
  Phys.\ Rev.\ D {\bf 73} (2006)125016.

\bibitem{dub06}
  S.~Dubni\v{c}ka, A.~Z.~Dubni\v{c}kov\'a, E.~A.~Kuraev, Phys. Rev.
  D {\bf 74} (2006) 034023.

\bibitem{dub07}
  S.~Dubni\v{c}ka, A.~Z.~Dubni\v{c}kov\'a, E.~A.~Kuraev, Phys. Rev.
  D {\bf 75} (2007) 057901.

 \bibitem{KEL}
 S.~R.~Kel'ner, Yad. Fiz. {\bf 5} (1967) 1092.

 \bibitem{NVY}
 M.~Nagy, M.~K.~Volkov, V.~L.~Yudichev, in: A.~Z.~Dubnickova,
 S.~Dubnicka, P.~Strizenec (Eds.), Proc. of Int. Conf. "Hadron
 Structure 2000", Stara Lesna, Slovak Republic, 2.-7.10.2000.,
 Comenius Univ., Bratislava, 2001, p.188.

 \bibitem{Yao}
 W.-M.~Yao et al, J. Phys. G {\bf 33} (2006) 1.



\end{thebibliography}
\end{document}